\documentclass[twocolumn]{aastex62}

\usepackage{amsmath}
\hypersetup{breaklinks=true,colorlinks=true,citecolor=blue,linkcolor=blue,urlcolor=blue}

\def\hst{\textit{HST}}

\shorttitle{The late-time lightcurve of ASASSN-14lp}
\shortauthors{Graur et al.}

\begin{document}

\title{Late-Time Observations of ASASSN-14lp Strengthen the Case for a Correlation between the Peak Luminosity of Type Ia Supernovae and the Shape of their Late-Time Light Curves}

\correspondingauthor{Or Graur}
\email{or.graur@cfa.harvard.edu}

\author{Or Graur}
\affiliation{Harvard-Smithsonian Center for Astrophysics, 60 Garden St., Cambridge, MA 02138, USA}
\affiliation{Department of Astrophysics, American Museum of Natural History, New York, NY 10024, USA}

\author{David R. Zurek}
\affiliation{Department of Astrophysics, American Museum of Natural History, New York, NY 10024, USA}

\author{Mihai Cara}
\affiliation{Space Telescope Science Institute, Baltimore, MD 21218, USA}

\author{Armin Rest}
\affiliation{Space Telescope Science Institute, Baltimore, MD 21218, USA}
\affiliation{Department of Physics and Astronomy, The Johns Hopkins University, Baltimore, MD 21218, USA}

\author{Ivo R. Seitenzahl}
\affiliation{School of Physical, Environmental, and Mathematical Sciences, University of New South Wales, Australian Defense Force Academy, Canberra, ACT 2600, Australia}
\affiliation{Research School of Astronomy and Astrophysics, The Australian National University, Cotter Road, Weston Creek, ACT 2611, Australia}

\author{Benjamin J. Shappee}
\altaffiliation{Hubble and Carnegie-Princeton Fellow}
\affiliation{The Observatories of the Carnegie Institution for Science, 813 Santa Barbara St., Pasadena, CA 91101, USA}
\affiliation{Institute for Astronomy, University of Hawai'i, 2680 Woodlawn Drive, Honolulu, HI 96822, USA}

\author{Michael M. Shara}
\affiliation{Department of Astrophysics, American Museum of Natural History, New York, NY 10024, USA}

\author{Adam G. Riess}
\affiliation{Space Telescope Science Institute, Baltimore, MD 21218, USA}
\affiliation{Department of Physics and Astronomy, The Johns Hopkins University, Baltimore, MD 21218, USA}


\begin{abstract}
 Late-time observations of Type Ia supernovae (SNe Ia), $>900$ days after explosion, have shown that this type of SN does not suffer an ``IR catastrophe'' at 500 days as previously predicted. Instead, several groups have observed a slow-down in the optical light curves of these SNe. A few reasons have been suggested for this slow-down, from a changing fraction of positrons reprocessed by the expanding ejecta, through a boost of energy from slow radioactive decay chains such as $^{57}$Co$\to^{57}$Fe, to atomic ``freeze-out.'' Discovering which of these (or some other) heating mechanisms is behind the slow-down will directly impact studies of SN Ia progenitors, explosion models, and nebular-stage physics. Recently, \citet{2018ApJ...859...79G} suggested a possible correlation between the shape of the late-time light curves of four SNe Ia and their stretch values, which are proxies for their intrinsic luminosities. Here, we present {\it Hubble Space Telescope} observations of the SN Ia ASASSN-14lp at $\sim 850$--960 days past maximum light. With a stretch of $s=1.15\pm0.05$, it is the most luminous normal SN Ia observed so far at these late times. We rule out contamination by light echoes and show that the late-time, optical light curve of ASASSN-14lp is flatter than that of previous SNe Ia observed at late times. This result is in line with---and strengthens---the \citet{2018ApJ...859...79G} correlation, but additional SNe are needed to verify it.
\end{abstract}



\section{Introduction}
\label{sec:intro}

The natures of the progenitors and explosion mechanism of Type Ia supernovae (SNe Ia) remain open questions (e.g., \citealt{2014ARA&A..52..107M}). Most SNe Ia are only studied for the first months of their lives, but over the last few years it has become clear that the very late nebular stage ($>500$ days after explosion) offers fresh clues to the progenitor and explosion physics of these SNe.
  
\citet{1980PhDT.........1A} predicted that $\sim500$ days after explosion, the cooling of the SN ejecta, which until this time proceeded through transitions in the optical, should switch to fine-structure transitions in the infrared (IR), resulting in a steep drop in the optical light curve --- the so-called ``IR catastrophe.'' To date, though, no such drop has been observed (e.g., \citealt{1997A&A...328..203C,2009A&A...505..265L,2014ApJ...796L..26K,2017ApJ...841...48S}). Instead, the optical decline of the light curve seems to slow down. Additional heating channels beyond the radioactive decay of $^{56}$Co may be required to explain this slow-down.
  
Several such channels have been suggested, such as slow radioactive decay chains (e.g., $^{57}$Co$\to^{57}$Fe with $t_{1/2}\approx272$ days) that begin to dominate the SN light at late times \citep{2009MNRAS.400..531S}, a changing fraction of positrons trapped by the ejecta as they continue to expand \citep{2017MNRAS.468.3798D,2017MNRAS.472.2534K}, or a ``freeze-out'' stage at which the recombination and cooling timescales become longer than the radioactive decay timescale, and the conversion of energy absorbed by the ejecta into emitted radiation is no longer instantaneous \citep{1993ApJ...408L..25F,2015ApJ...814L...2F}.
    
Thus the shape of a SN Ia light curve at $>500$ days encodes information about the progenitor of the SN, how it exploded, and the physics of its nebular phase. \citet{2016ApJ...819...31G} first showed that the light curve of SN 2012cg was consistent with a combination of $^{56}$Co and $^{57}$Co radioactive decays. The same was shown by \citet{2017ApJ...841...48S} for SN 2011fe and by \citet{2018ApJ...852...89Y} for SN 2014J. According to \citet{2017ApJ...841...48S}, the late-time photometry of SN 2011fe was precise enough to rule out the single-degenerate progenitor scenario \citep{Whelan1973}, but \citet{2017MNRAS.468.3798D} and \citet{2017MNRAS.472.2534K} showed that similar data could also be fit with models that assumed either freeze-out or a changing fraction of positron trapping. 

A single SN, then, does not have sufficient power to rule out any of the existing models. \citet[G18]{2018ApJ...859...79G} showed a tentative correlation between the stretch of SNe Ia (a proxy for their intrinsic luminosities, as codified by the \citealt{1993ApJ...413L.105P} relation) and the shape of their late-time light curves. Such a correlation has the potential to constrain all of the nebular, explosion, and progenitor models described above. However, it is based on only four SNe -- the only ones followed out to sufficiently late times. Recently, \citet{2018ApJ...857...88J} cast doubt on this correlation, based on a single observation of SN 2013aa at $\approx 1500$ days. 

Here, we present \textit{Hubble Space Telescope} (\hst) observations of ASASSN-14lp between $\sim 850$ and $\sim 960$ days past maximum light. In Section~\ref{sec:data}, we photometer the SN and construct its pseudo-bolometric light curve. In Section~\ref{sec:analysis}, we show that the SN is not contaminated by light echoes, and compare it to the four SNe studied by G18. Although the data for ASASSN-14lp are not enough to fit the various heating mechanisms discussed above, they indicate the continuation of the correlation claimed by G18, namely that more luminous SNe Ia (with higher stretch values) have flatter late-time light curves. We offer concluding remarks in Section~\ref{sec:discuss}.

\begin{figure*}
 \centering
 \includegraphics[width=0.97\textwidth]{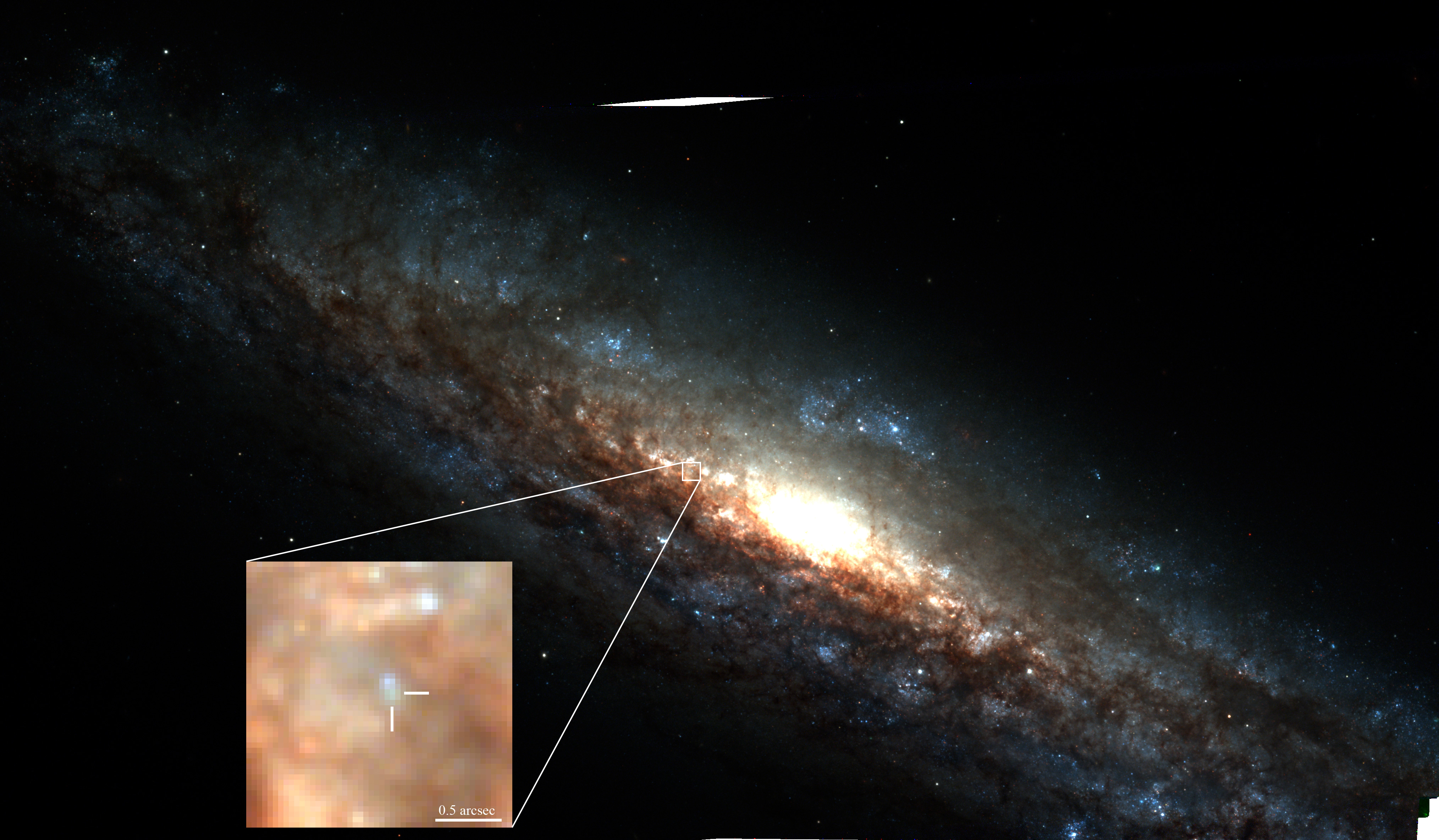}
 \caption{\textit{HST} color composite of ASASSN-14lp near the core of NGC 4666. The image is composed of $F814W$ (red), $F555W$ (green), and $F438W$ (blue) images from all visits to NGC 4666 in program GO--14611. The location of the SN is marked by a $2^{\prime\prime}\times2^{\prime\prime}$ white box, which is blown up in the inset, where ASASSN-14lp is identified by a white reticle. North is up and East is to the left, tilted by $21^{\circ}$ counter-clockwise.}
 \label{fig:host}
\end{figure*}


\section{Observations and Photometry}
\label{sec:data}

ASASSN-14lp was discovered by the All-Sky Automated Survey for Supernovae (ASAS-SN; \citealt{2014ApJ...788...48S}) on 2014 December 9.6 \citep{2016ApJ...826..144S} in the nearby spiral galaxy NGC 4666. \citet{2016ApJ...826..144S} used the SN to derive a luminosity distance to the galaxy of $14.7 \pm 1.5$ Mpc, which we adopt here. They also measured a broad light curve with a $\Delta m_{15}(B)$ value of $0.80 \pm 0.05$, making it a luminous but still normal SN Ia. This $\Delta m_{15}(B)$ converts to a SiFTO stretch value of $s=1.15 \pm 0.05$ \citep{2008ApJ...681..482C}.

We imaged ASASSN-14lp ($\alpha=12^{\rm h}45^{\rm m}09.10^{\rm s}$, $\delta=-00^{\circ}27^{\prime}32.5^{\prime\prime}$) with the \hst\ Wide Field Camera 3 (WFC3) wide-band filters {\it F438W}, {\it F555W}, {\it F625W}, and {\it F814W} under \hst\ program GO--14611 (PI: Graur) on five separate occasions between 2017 April 21 and 2017 August 7. At these times, the SN was $\sim 850$--$960$ days past maximum light. The observation log is presented in Table~\ref{table:mags_14lp}. These observations can be found in the Mikulski Archive for Space Telescopes (MAST) at \href{http://archive.stsci.edu/doi/resolve/resolve.html?doi=10.17909/T9240M}{DOI:10.17909/T9240M}. 

In Figure~\ref{fig:host}, we show a color composite of NGC 4666 and ASASSN-14lp. Throughout this work, the phases we cite are calculated relative to $B$-band maximum light, which occurred on 2014 December 24 (JD $2457015.82 \pm 0.03$; \citealt{2016ApJ...826..144S}).

\begin{figure*}
 \centering
 \includegraphics[width=0.95\textwidth]{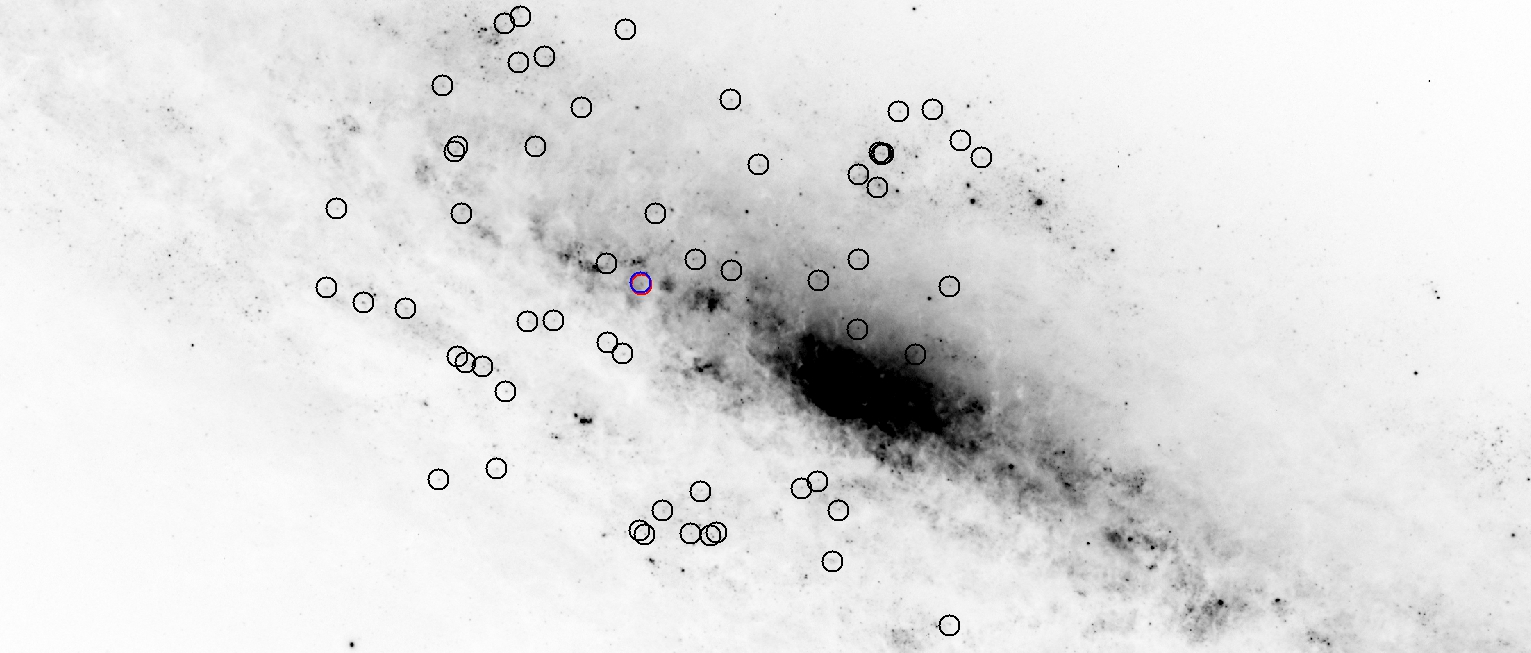}
 \begin{tabular}{cc}
  \includegraphics[width=0.475\textwidth]{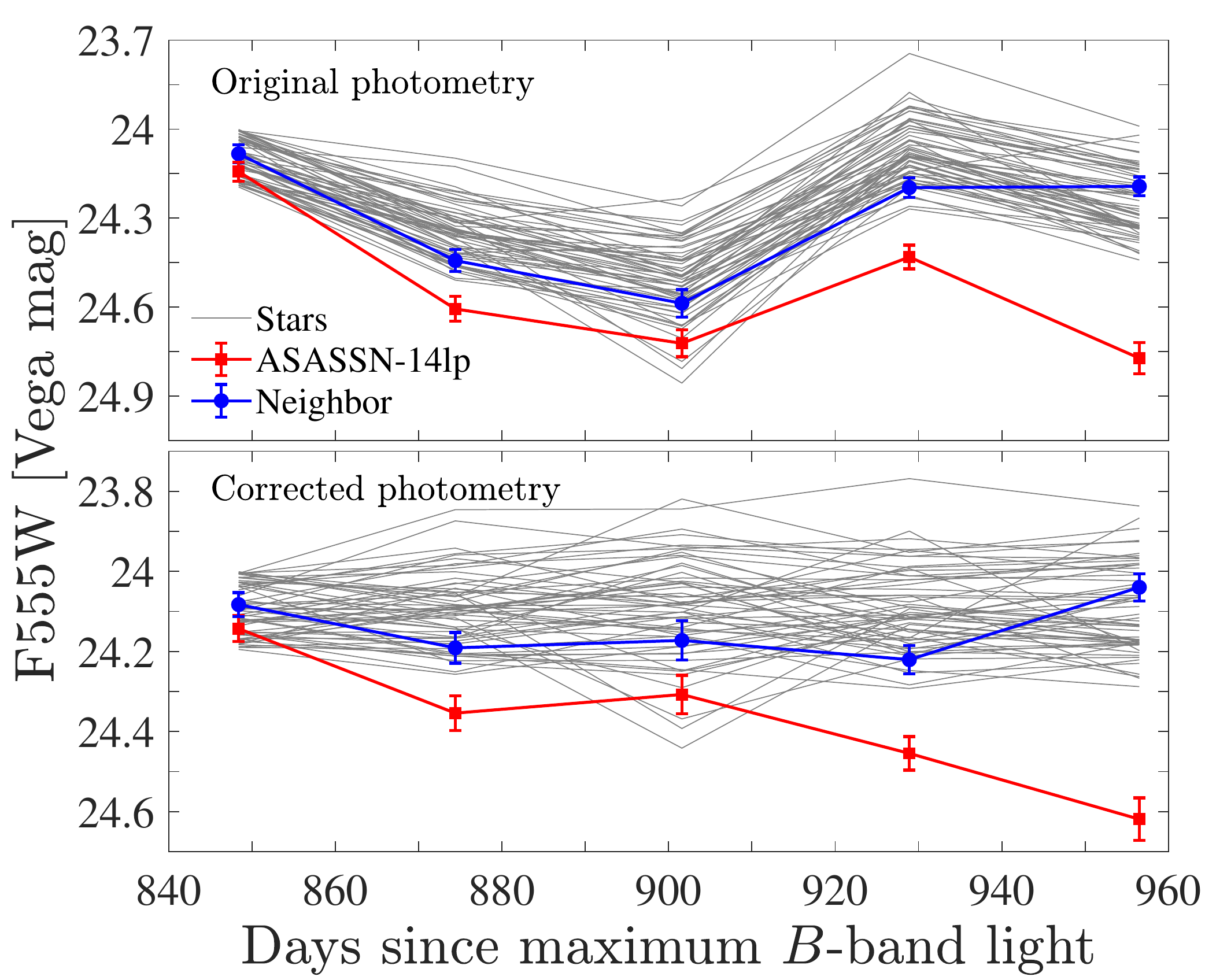} &
  \includegraphics[width=0.475\textwidth]{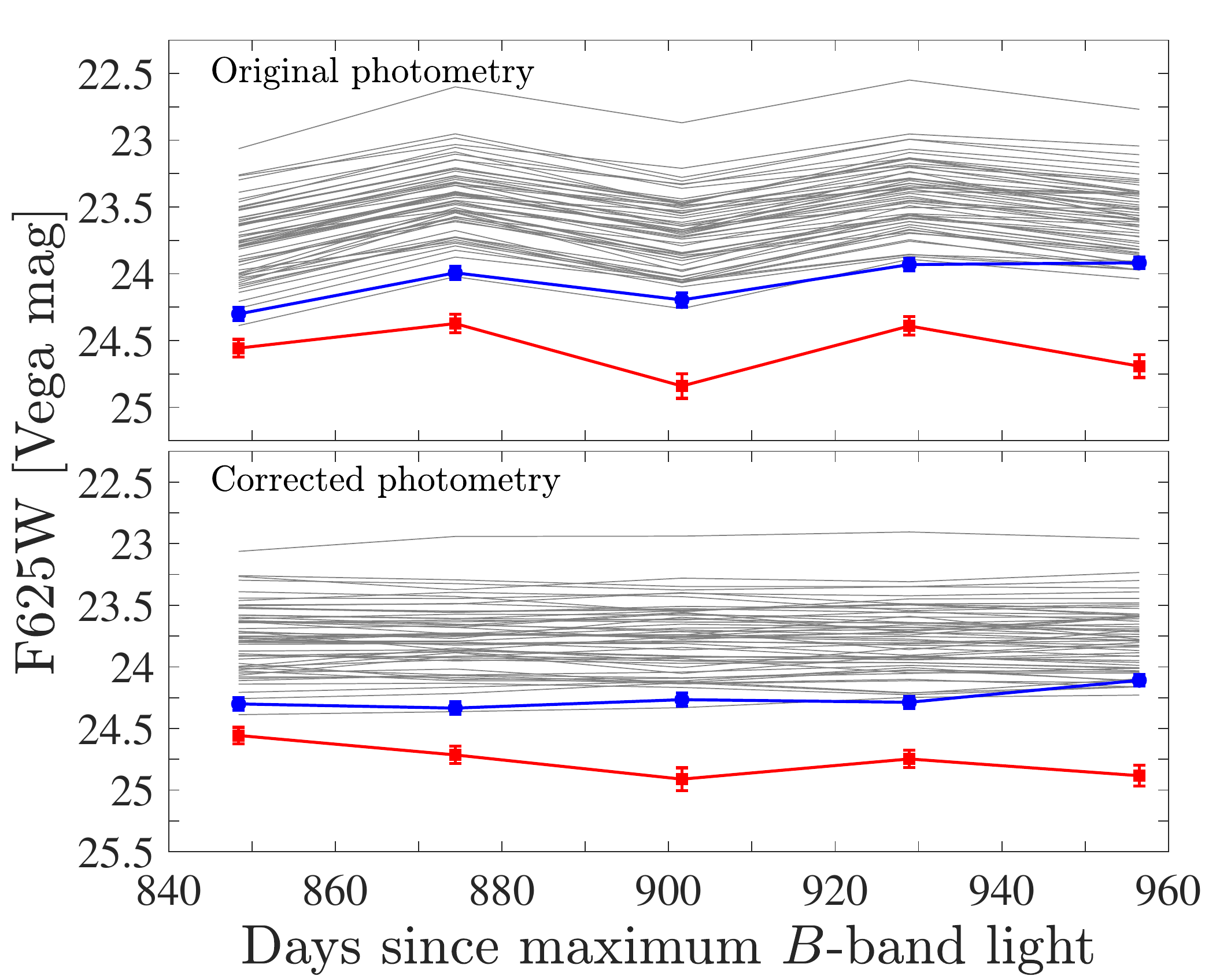} \\
  \includegraphics[width=0.475\textwidth]{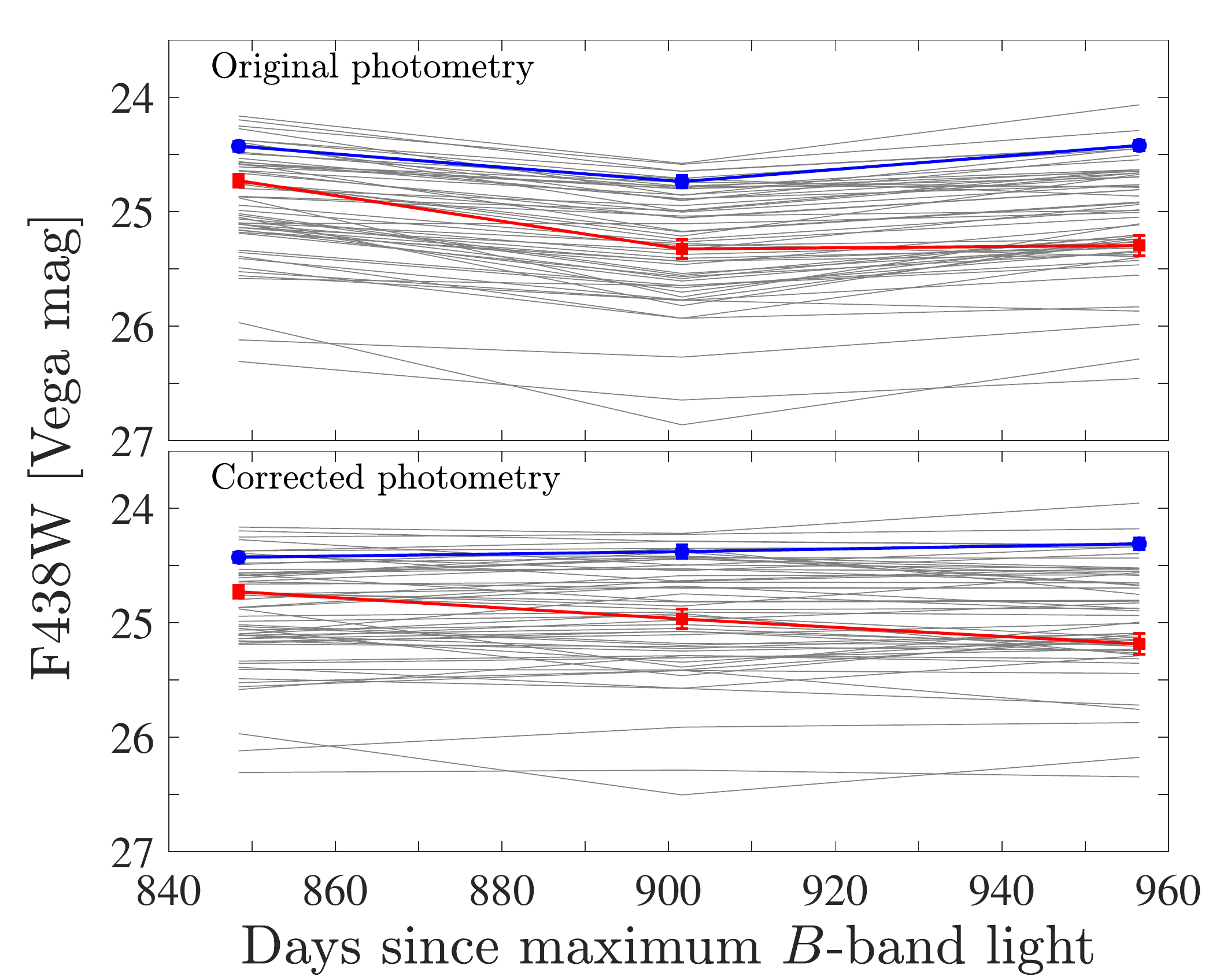} &
  \includegraphics[width=0.475\textwidth]{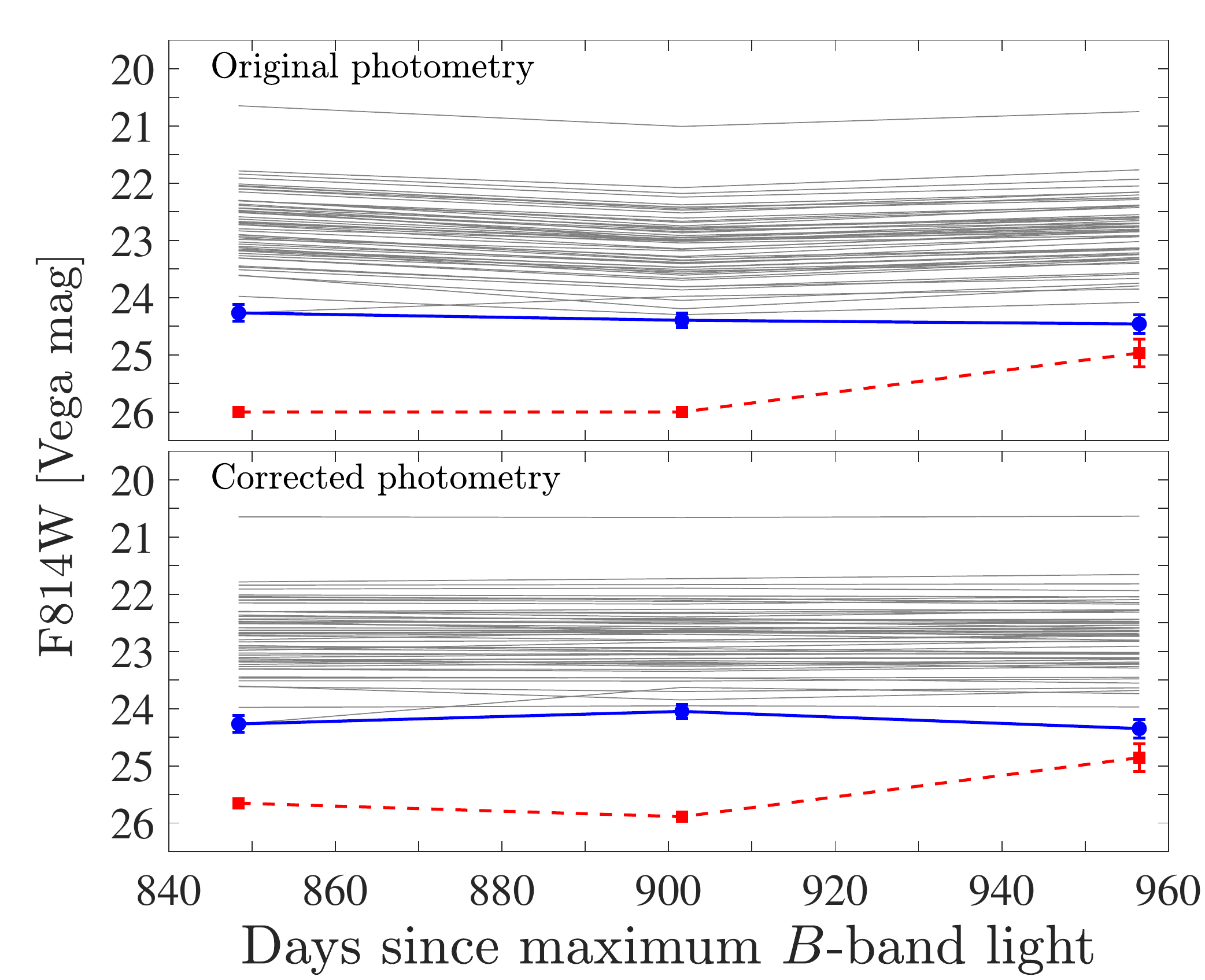} \\
 \end{tabular}
 \caption{Upper panel: \textit{F555W} image of NGC 4666 showing the 58 point sources (circled in black) used to correct the photometry of ASASSN-14lp. The SN is circled in red, and its neighboring point source---in blue. Center and bottom panels: the upper and lower half of each panel shows, respectively, the original and corrected photometry of the 58 point sources (gray lines). The SN and its neighboring point source are shown in red squares and blue circles, respectively. Correcting the photometry of the SN by using either the mean differences in magnitude between the first and each successive visit of all 58 point sources, or only those of the neighboring point source, provide consistent results.}
 \label{fig:magstars}
\end{figure*}

\subsection{PSF-fitting photometry}
\label{subsec:phot}

We used the {\sc Tweakreg} and {\sc AstroDrizzle} tasks included in the {\sc DrizzlePac} Python package\footnote{\url{http://drizzlepac.stsci.edu/}} to align the \hst\ images and remove cosmic rays and bad pixels. Next, we performed point-spread-function (PSF) fitting photometry using {\sc Dolphot}\footnote{\url{http://americano.dolphinsim.com/dolphot/}} \citep{2000PASP..112.1383D} on the {\sc flc} files produced by the \hst\ WFC3 pipeline, which are corrected for charge transfer efficiency effects. The resulting photometry, in Vega mags, is presented in Table~\ref{table:mags_14lp}.

Although the images were reduced in the same manner as in G18, point sources across the image fluctuated in brightness between visits in a systematic manner, likely due to breathing of the PSF (e.g., \citealt{2012wfc..rept...14D,2013wfc..rept...11S,2015wfc..rept....8A}). To correct for this effect, shown in Figure~\ref{fig:magstars}, we selected 58 point sources within a radius of $20^{\prime\prime}$ around the SN. These point sources were required to be within the magnitude range $24.0<F555W<24.2$~mag, have a {\sc Dolphot} sharpness index in the range $-0.3$--$0.3$ (nominal for stars), and fade between the first and second visits (a small fraction of point sources brightened between these visits, rather than declined).

Assuming that these point sources should not vary in brightness across epochs, we measured the differences between the first and all succeeding visits and calculated the mean difference and its standard deviation (divided by the square root of the number of sources) in each epoch. We also calculated what the photometry of the SN would look like if we corrected it according to the neighboring point source $\approx 0.1^{\prime\prime}$ to the NE of the SN, and found that the results were consistent within the error bars. The latter are composed of the uncertainties of the original photometry and mean corrections, added in quadrature. Both sets of corrected photometry are listed in Table~\ref{table:mags_14lp}, but from here on we adopt the mean-corrected set.

ASASSN-14lp is only marginally detected in the final {\it F814W} visit. In the first and third visits, we adopt upper limits of $>26.0$ mag (before correction), which represent the expected brightness of a point source with a signal-to-noise (S/N) ratio of 5.

A potential source of systematic uncertainty in our photometry is crowding of the SN by other point sources. This was not taken into account by G18, since in that paper the authors only fit the \emph{shape} of the light curve, and thus a constant addition to the SN luminosity would not have made a difference. In Section~\ref{sec:analysis}, however, such an addition could affect our results. We test for crowding of the SN by examining the ``crowding'' parameter measured by {\sc Dolphot}. This parameter, given in magnitudes, indicates how much brighter a point source would be if other, nearby sources were not fit simultaneously along with it. Higher values indicate a denser field of sources. SNe 2012cg and 2015F, which exploded in visually sparse regions, had crowding values of $0.1$--$0.2$ mag and $0$ mag, respectively. The photometry of ASASSN-14lp, in all filters, had similar values of $0.1$--$0.2$ mag. We thus conclude that our photometry is not strongly impacted by crowding.

\subsection{Pseudo-bolometric light curve}
\label{subsec:pbl}

We follow \citet{2017ApJ...841...48S} and G18 and construct a pseudo-bolometric light curve from our \textit{HST} observations. For a complete description of this process, see G18; here, we only describe any differences to the G18 process required for ASASSN-14lp.
  
All magnitudes are corrected for Galactic and host-galaxy extinction. The Galactic line-of-sight extinctions in $F438W$, $F555W$, $F625W$, and $F814W$ are $0.105$, $0.068$, $0.057$, and $0.042$ mag, respectively \citep{2011ApJ...737..103S}. Using the \citet{1989ApJ...345..245C} reddening law, a measured $E(B-V)=0.33 \pm 0.06$ mag \citep{2016ApJ...826..144S}, and assuming $R_V=3.1$, we estimate host-galaxy extinctions of $1.383$, $1.065$, $0.888$, and $0.608$ mag in the same filters.

To account for the non-detections of the SN in {\it F814W}, we used the upper limits measured in the first and third visit, with inflated uncertainties of 0.5 mag. The resultant luminosity in {\it F814W} contributes only $\sim2$--8\% to the total pseudo-bolometric light curve between the first and last visit, respectively.


\section{Analysis}
\label{sec:analysis}

In this section, we use the corrected photometry and pseudo-bolometric light curve derived in Sections~\ref{subsec:phot} and \ref{subsec:pbl} to show that ASASSN-14lp is not contaminated by light echoes (Section~\ref{subsec:le}) and that it continues the trend, first shown by G18, that more luminous SNe Ia have flatter light curves at late times (Section~\ref{subsec:correlation}).

\subsection{No contamination by light echoes}
\label{subsec:le}

Light echoes, produced when the light of the SN is reflected off nearby dust sheets, have been shown to flatten the light curves of some SNe Ia and shift their colors blueward beginning at $\sim 500$ days \citep{1994ApJ...434L..19S,1999ApJ...523..585S,2006ApJ...652..512Q,2001ApJ...549L.215C,2008ApJ...677.1060W,2015ApJ...805...71D,2015ApJ...804L..37C}. As in G18, we test whether ASASSN-14lp is contaminated by a light echo by comparing its colors to those of SN 2011fe, which is clear of such contamination \citep{2015MNRAS.454.1948G,2017ApJ...841...48S}. 

\begin{figure}
 \centering
 \includegraphics[width=0.47\textwidth]{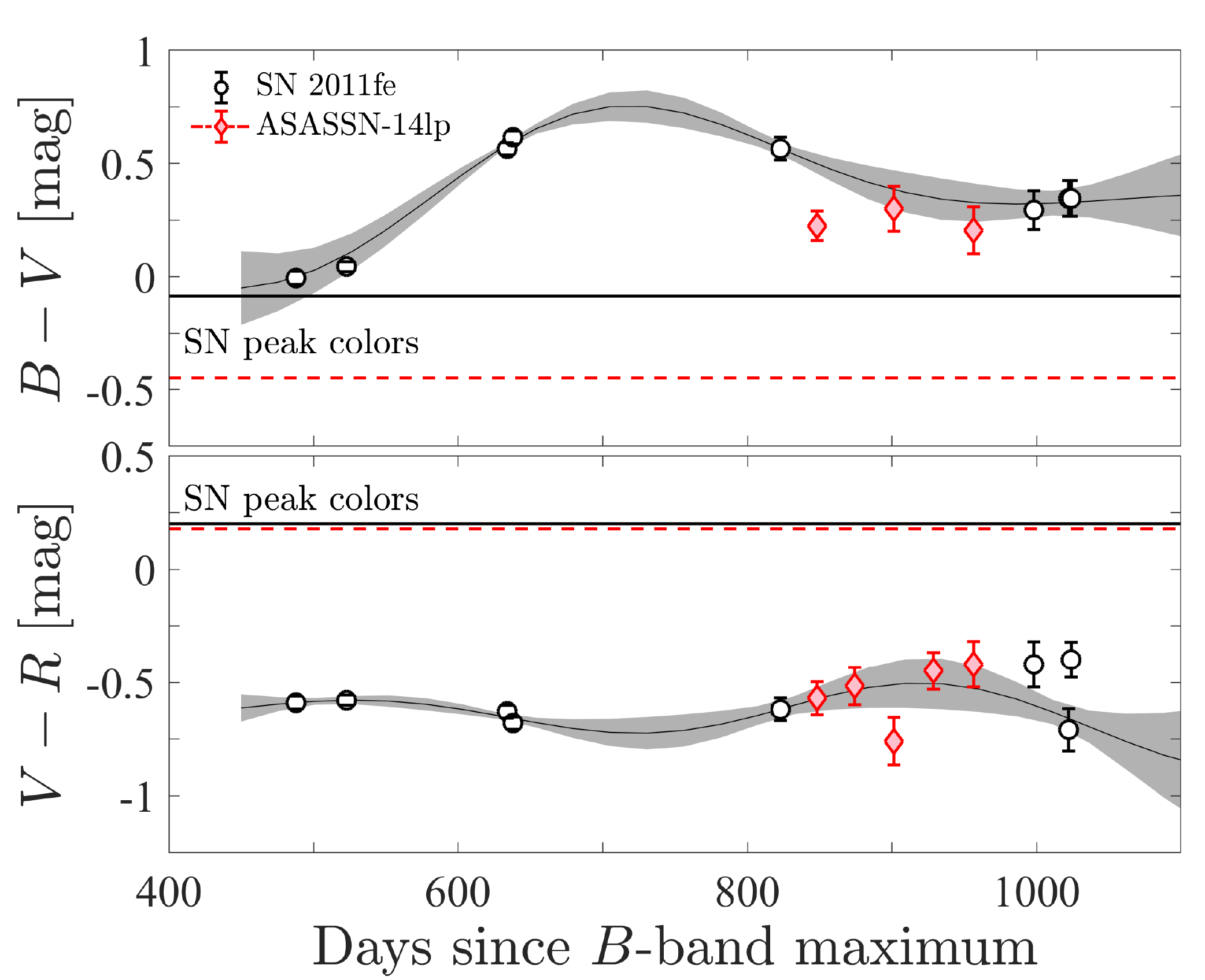}
 \caption{A comparison between the $B-V$ (top) and $V-R$ (bottom) colors of ASASSN-14lp (red diamonds) and SN 2011fe (black circles). In each panel, the dashed, and solid curves represent the $B-V$ or $V-R$ colors of these SNe, respectively, at $B$-band maximum light. The gray shaded bands connecting the colors of SN 2011fe are Gaussian Process regressions; the width of the band represents the 68\% uncertainty of the fit. The colors of ASASSN-14lp are broadly consistent with those of SN 2011fe. Importantly, both SNe are redder in $B-V$ and bluer in $V-R$ than their colors at peak light. A light echo, which shifts the spectrum blueward, cannot account for such an effect.}
 \label{fig:color}
\end{figure}

Figure~\ref{fig:color} shows the $B-V$ and $V-R$ colors of ASASSN-14lp, compared to those of SN 2011fe \citep{2017ApJ...841...48S}. We also compare these colors to the colors of the SNe at peak (as measured by \citealt{2013NewA...20...30M} and \citealt{2016ApJ...826..144S} for SNe 2011fe and ASASSN-14lp, respectively). As in G18, all magnitudes have been corrected for Galactic and host-galaxy extinction, and converted between the original filter sets and the Johnson-Cousins $BVR$ filters used in this test. 

The colors of ASASSN-14lp are consistent with those of SN 2011fe. Moreover, compared to its peak colors, ASASSN-14lp is redder in $B-V$, but bluer in $V-R$. Light echoes preferentially scatter light to bluer wavelengths, making the SNe appear to be bluer than they were at peak. Thus, a light echo cannot account for both of the above effects. We conclude that ASASSN-14lp is not contaminated by light echoes, at least within the phase range probed here.

\subsection{The shape of the late-time light curve}
\label{subsec:correlation}

\begin{figure*}
 \centering
 \begin{tabular}{cc}
  \includegraphics[width=0.47\textwidth]{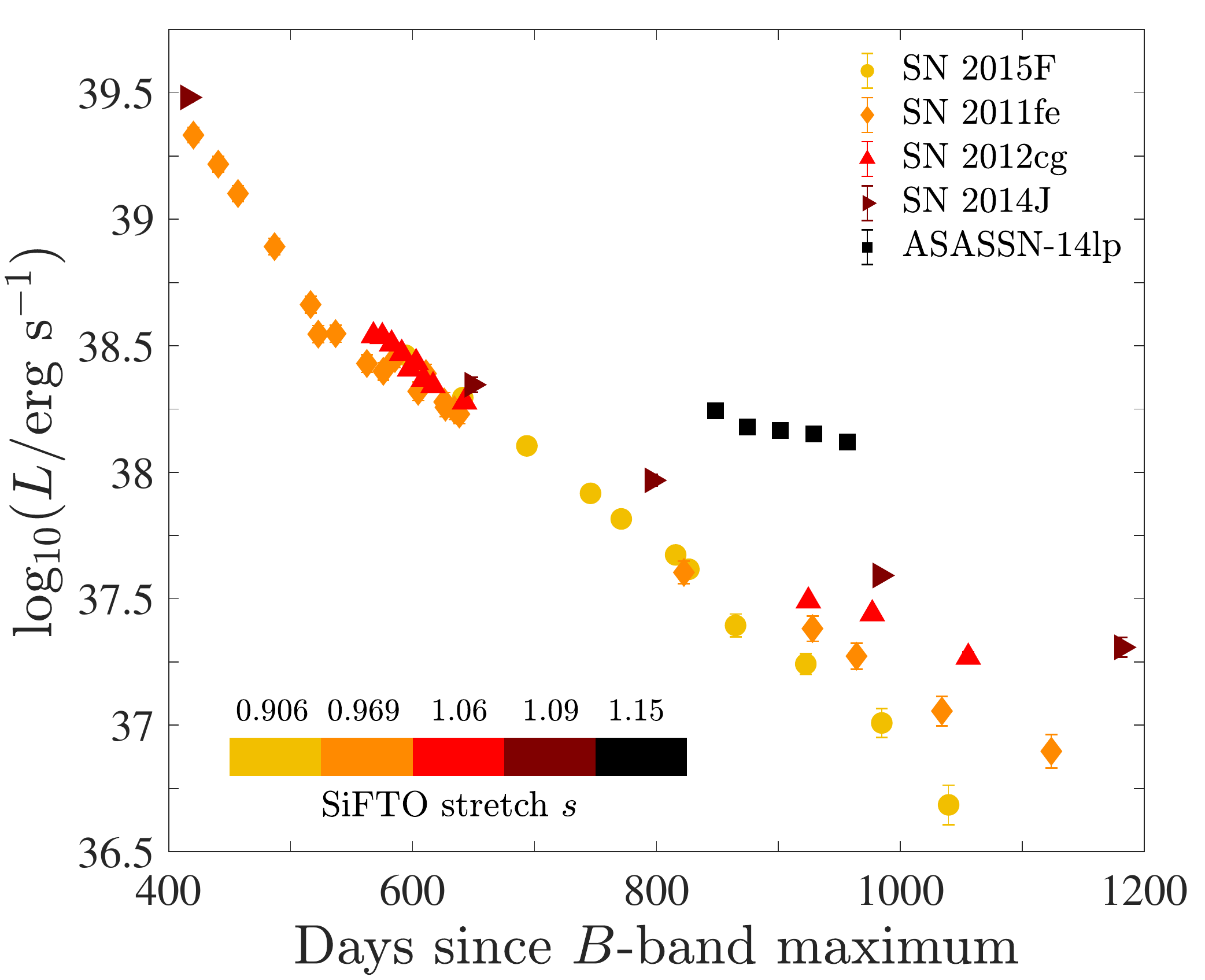} & \includegraphics[width=0.47\textwidth]{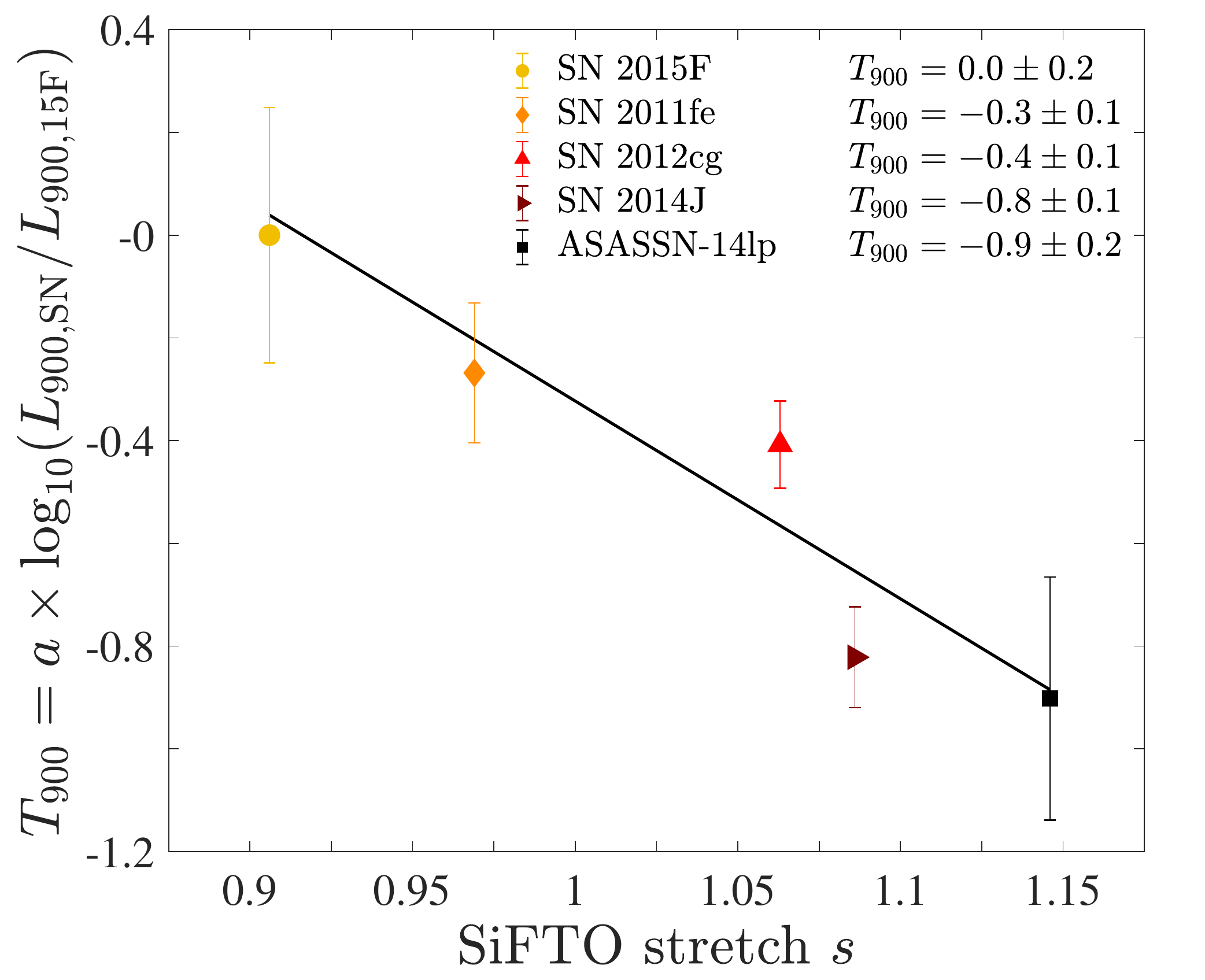} \\
 \end{tabular}
 \caption{Left panel: the pseudo-bolometric light curve of ASASSN-14lp (black squares), compared to those of previous SNe Ia. The colormap is chosen to show that SNe with higher stretch values (darker hues) have flatter light curves at late times, at least between 800 and 1000 days. The single point measured by \citet{2018ApJ...857...88J} for SN 2013aa at $\approx 1500$ days is not shown here. Right panel: $T_{900}$, which accounts for both the convergence of the SN Ia light curves at $\sim 500$--700 days and their divergence at $>800$ days, as a function of stretch. A likelihood ratio test finds a $>3\sigma$ correlation (solid curve). The data have a Pearson correlation coefficient of $\rho=-0.95$ with a $p$-value of $0.013$.}
 \label{fig:pbl}
\end{figure*}

The pseudo-bolometric light curve of ASASSN-14lp, shown in the left panel of Figure~\ref{fig:pbl}, shows that this SN did not go through an IR catastrophe at $\sim 500$ days, as predicted by \citet{1980PhDT.........1A}. If that were the case, we would not have detected the SN in our images at all.

Previous works have used pseudo-bolometric light curves to fit for the mass ratio of $^{57}$Co$/^{56}$Co, the time at which freeze-out sets in, or the varying fraction of positrons reprocessed by the ejecta. G18 also suggested a model-independent method to compare between the late-time light curves of SNe Ia, by comparing their luminosities at 600 and 900 days. Unfortunately, because the late-time observations of ASASSN-14lp only cover the phase range $\sim 850$--960 days, the data cannot place strong constraints on the models above. 

\floattable
\begin{deluxetable}{lCCcCCCCC}
 \tablecaption{Observation log for ASASSN-14lp. \label{table:mags_14lp}}
 \tablehead{
 \colhead{Date} & \colhead{MJD}    & \colhead{Phase}  & \colhead{Filter} & \colhead{Exposure} & \colhead{Measured}   & \colhead{Mean-corrected} & \colhead{Neighbor-corrected} & \colhead{Luminosity} \\
 \colhead{}     & \colhead{}       & \colhead{}       & \colhead{}       & \colhead{Time}     & \colhead{Magnitude}  & \colhead{Magnitude}      & \colhead{Magnitude}          & \colhead{} \\
 \colhead{}     & \colhead{(days)} & \colhead{(days)} & \colhead{}       & \colhead{(s)}      & \colhead{(Vega mag)} & \colhead{(Vega mag)}     & \colhead{(Vega mag)}         & \colhead{${\rm log}(L/{\rm erg~s^{-1}})$} 
 }
 \decimals
 \startdata
  2017 Apr. 21.2 & 57864.2 & 848.4 & {\it F438W} & 1140   & 24.73 \pm 0.06 & 24.73 \pm 0.06 & 24.73 \pm 0.06 & 37.73 \pm 0.02 \\
  2017 Apr. 21.2 & 57864.2 & 848.4 & {\it F555W} & 1200   & 24.14 \pm 0.03 & 24.14 \pm 0.03 & 24.14 \pm 0.03 & 38.00 \pm 0.01 \\
  2017 Apr. 21.2 & 57864.2 & 848.4 & {\it F625W} & 1134   & 24.56 \pm 0.07 & 24.56 \pm 0.07 & 24.56 \pm 0.07 & 37.52 \pm 0.03 \\
  2017 Apr. 21.2 & 57864.2 & 848.4 & {\it F814W} & 1152   & >26.0          & > 25.7         & > 25.9         & <36.6          \\
  2017 Apr. 21.2 & 57864.2 & 848.4 & Optical     & \cdots & \cdots         & \cdots         & \cdots         & 38.24 \pm 0.02 \\ 
  2017 May  17.2 & 57890.2 & 874.4 & {\it F555W} & 1143   & 24.61 \pm 0.04 & 24.35 \pm 0.04 & 24.25 \pm 0.04 & 37.91 \pm 0.02 \\
  2017 May  17.2 & 57890.2 & 874.4 & {\it F625W} & 1140   & 24.37 \pm 0.07 & 24.71 \pm 0.07 & 24.68 \pm 0.07 & 37.45 \pm 0.03 \\
  2017 May  17.2 & 57890.2 & 874.4 & Optical     & \cdots & \cdots         & \cdots         & \cdots         & 38.18 \pm 0.02 \\
  2017 Jun. 13.4 & 57917.4 & 901.6 & {\it F438W} & 1140   & 25.33 \pm 0.08 & 24.97 \pm 0.09 & 25.02 \pm 0.09 & 37.64 \pm 0.03 \\
  2017 Jun. 13.4 & 57917.4 & 901.6 & {\it F555W} & 1200   & 24.72 \pm 0.05 & 24.31 \pm 0.05 & 24.22 \pm 0.05 & 37.93 \pm 0.02 \\
  2017 Jun. 13.4 & 57917.4 & 901.6 & {\it F625W} & 1134   & 24.84 \pm 0.09 & 24.91 \pm 0.09 & 24.95 \pm 0.09 & 37.38 \pm 0.04 \\
  2017 Jun. 13.4 & 57917.4 & 901.6 & {\it F814W} & 1152   & >26.0          & > 25.9         & >25.8          & <36.6          \\
  2017 Jun. 13.4 & 57917.4 & 901.6 & Optical     & \cdots & \cdots         & \cdots         & \cdots         & 38.17 \pm 0.02 \\
  2017 Jul. 10.7 & 57944.7 & 928.9 & {\it F555W} & 1143   & 24.43 \pm 0.04 & 24.45 \pm 0.04 & 24.32 \pm 0.04 & 37.87 \pm 0.02 \\
  2017 Jul. 10.7 & 57944.7 & 928.9 & {\it F625W} & 1140   & 24.39 \pm 0.07 & 24.75 \pm 0.07 & 24.76 \pm 0.07 & 37.44 \pm 0.03 \\
  2017 Jul. 10.7 & 57944.7 & 928.9 & Optical     & \cdots & \cdots         & \cdots         & \cdots         & 38.15 \pm 0.02 \\
  2017 Aug.  7.3 & 57972.3 & 956.5 & {\it F438W} & 1140   & 25.30 \pm 0.09 & 25.18 \pm 0.09 & 25.30 \pm 0.09 & 37.55 \pm 0.04 \\
  2017 Aug.  7.3 & 57972.3 & 956.5 & {\it F555W} & 1200   & 24.77 \pm 0.05 & 24.62 \pm 0.05 & 24.66 \pm 0.05 & 37.81 \pm 0.02 \\
  2017 Aug.  7.3 & 57972.3 & 956.5 & {\it F625W} & 1134   & 24.69 \pm 0.09 & 24.88 \pm 0.09 & 25.08 \pm 0.09 & 37.39 \pm 0.03 \\
  2017 Aug.  7.3 & 57972.3 & 956.5 & {\it F814W} & 1152   & 25.0  \pm 0.2  & 24.9  \pm 0.2  & 24.8  \pm 0.2  & 37.0  \pm 0.1  \\
  2017 Aug.  7.3 & 57972.3 & 956.5 & Optical     & \cdots & \cdots         & \cdots         & \cdots         & 38.12 \pm 0.03 \\
 \enddata
 \textbf{Note.} All photometry is measured using PSF-fitting photometry with {\sc Dolphot}. ``Optical'' refers to the pseudo-bolometric luminosities, in the wavelength range $3500$--$10000$~\AA, derived in Section~\ref{subsec:pbl}. Upper limits represent the median measured magnitude of point sources with a S/N of 5. Throughout this work, we adopt the ``mean-corrected'' magnitudes. All observations can be found in MAST at \href{http://archive.stsci.edu/doi/resolve/resolve.html?doi=10.17909/T9240M}{DOI:10.17909/T9240M}.
\end{deluxetable}

Instead, we limit this section to a simple test of the correlation claimed by G18. According to that work, more luminous SNe Ia, as indicated by their stretch value, should exhibit a flatter light curve at late times, beginning at $\sim 800$ days. When fit with the heating mechanisms described in Section~\ref{sec:intro}, SNe Ia with flatter late-time light curves should have either higher $^{57}$Co$/^{56}$Co  mass ratios or earlier freeze-out times. 

Figure~\ref{fig:pbl} shows that ASASSN-14lp continues the trend claimed by G18. The left panel is color-coded according to the stretch values of the SNe, with SNe with higher values (i.e., more luminous) shown in darker hues. 

To fully capture the information encoded in the late-time light curves, G18 parameterized them according to the ratio of their luminosities at 600 and 900 days, which they termed $\Delta L_{900}$. Because we did not observe ASASSN-14lp at 600 days, we define here a new parameter, $T_{900}$, so that
\begin{equation}
 T_{900} = a \times {\rm log_{10}}\left(\frac{L_{900,{\rm SN}}}{L_{900,{\rm 15F}}}\right),
\end{equation}
where $a$ is the slope of a linear fit of the form ${\rm log}_{10}(L/{\rm erg~s^{-1}}) = a \times (t/10^3~{\rm days})+b$, and $t$ is limited to $>800$ days.\footnote{Except for SN 2014J, for which we include the observation at $796.4$ days.} In this way, $T_{900}$ captures the same properties of the late-time light curves for which $\Delta L_{900}$ was designed: (1) the interesting fact that the light curves of all SNe Ia observed so far at these phases broadly overlap at $\sim 500$--$700$ days (as seen in Figure~\ref{fig:pbl}, at $600$ days the SNe span a luminosity range of only $\approx 0.2$ dex, with error bars of $0.004$--$0.04$ dex); and (2) the slopes of their light curves at $>800$ days, when the SNe continue to fade at significantly different rates (at $900$ days, the luminosity span increases to $\approx 0.9$ dex, or $\approx 0.5$ dex if ASASSN-14lp is excluded). We compare the SNe to SN 2015F because G18 showed that its light curve was consistent with the pure radioactive decay of $^{56}$Co.

In the right panel of Figure~\ref{fig:pbl}, we compare the $T_{900}$ value of ASASSN-14lp to those of the SNe Ia tested by G18. A likelihood ratio test (see \citealt{2017ApJ...837..120G}) reveals that a linear fit (a correlation) to the latter measurements is preferable over a constant (no correlation) at a significance of $>3\sigma$. These data also have a Pearson's correlation coefficient of $\rho=-0.95$ with a $p$-value of $0.013$. If, instead of the mean-corrected photometry of ASASSN-14lp used throughout this section, we substitute the neighbor-corrected photometry, the $T_{900}$ value for ASASSN-14lp drops from $-0.9 \pm 0.2$ to $-1.1 \pm 0.1$. These values are consistent with each other within the error bars and have no effect on the results of the likelihood ratio test or Pearson's correlation coefficient. 

Above, we have assumed that the light curve of ASASSN-14lp would have converged with those of the other SNe Ia if we had obtained observations at $500$--$700$ days. A simpler test, which only uses the slopes of the light curves as fit at $>800$ days, provides similar results: a $>6\sigma$ significance according to a likelihood ratio test (which drops to $>2\sigma$ if SN 2015F is excluded), and a Pearson coefficient of $0.88$ with a $p$-value of $0.052$. Thus, there is still the suggestion of a correlation.

Although the results of the statistical tests used here are formally significant, we prefer to be cautious and repeat the conclusion arrived at by G18 that the data only \emph{suggest} a correlation. First, our analysis is based on only five SNe. Second, the only independent test of this correlation performed so far conflicts with our results. By comparing observations of SN 2013aa at $<400$ days with a single late-time observation at $\approx 1500$ days, \citet{2018ApJ...857...88J} claimed that the SN significantly fell off the correlation. Because SN 2013aa was not observed in the phase range probed here, we did not include it in our analysis. It remains to be seen whether future SNe Ia observed in this phase range will conform to the G18 correlation or not.


\section{Conclusions}
\label{sec:discuss}

We used \hst\ to observe the luminous, normal SN Ia ASASSN-14lp in the wavelength range $\sim3500$--$10000$~\AA\ when the SN was $\approx 850$--$960$ days past maximum light. As with previous SNe Ia observed at these late times, there is no evidence of an ``IR catastrophe.'' Instead, the light curve of ASASSN-14lp is seen to flatten out. We used the colors of this SN to rule out contamination by a light echo, so we surmise that the slow-down of the light curve should be due to one or more heating mechanisms that kick in at late times and dominate over the ongoing nuclear decay of $^{56}$Co.

We set out to test the claim made by G18 that more luminous SNe, as described by their stretch values, have flatter late-time light curves. ASASSN-14lp seems to follow this trend. However, the phase range of the observations is too short to constrain the theoretical heating models tested in previous works or to use the $\Delta L_{900}$ parameter designed by G18. Instead, we parameterize the late-time light curves with a new parameter, $T_{900}$, which accounts for both the convergence of the light curves at $\sim 500$--$700$ days, and their diverging slopes at $>800$ days. 

The $T_{900}$ values of ASASSN-14lp and the four SNe Ia tested by G18 are formally correlated according to the statistical tests used here. However, because our analysis is based on only five objects, we refrain from claiming a high statistical significance. To conduct a definitive test of the possible correlation between the luminosities of the SNe Ia and the shapes of their late-time light curves, a larger set of SNe Ia needs to be observed with \hst\ in a self-consistent experiment, i.e., by using the same observational setup and covering the entire phase range of $600$--$1000$ days.

Finally, we encourage independent tests of the G18 correlation. \citet{2018ApJ...857...88J} conducted the first such test. Comparing between one late-time observation of SN 2013aa at $\approx 1500$ days and earlier observations at $<400$ days, they claimed that the SN significantly fell off the correlation. Because this SN had no data in the phase range probed here, we could not analyze it together with ASASSN-14lp. We also caution that multiple observations are required at $>800$ days to properly describe the shape of the late-time light curve. Still, SN 2013aa underscores the need for more SNe Ia to properly test the claim made by G18.


\section*{Acknowledgments}

We thank Linda Dressel, Weston Eck, Mario Gennaro, and Blair Porterfield at the Space Telescope Science Institute (STScI) for shepherding Program GO--14611, and Daniel Eisenstein for helpful discussions and comments. OG was supported by NASA through \hst-GO--14611. IRS acknowledges funding from the Australian Research Council under grant FT160100028. BS was partially supported by NASA through \hst-GO--14166 and 14678 and Hubble Fellowship grant HF-51348.001. This work is based on data obtained with the NASA/ESA {\it Hubble Space Telescope}, all of which was obtained from MAST. Support for Programs GO--14166, GO--14611, and GO--14678 was provided by NASA through grants from STScI, which is operated by the Association of Universities for Research in Astronomy, Incorporated, under NASA contract NAS5-26555. Support for MAST for non-\hst\ data is provided by the NASA Office of Space Science via grant NNX09AF08G and by other grants and contracts. This research has made use of NASA's Astrophysics Data System and the NASA/IPAC Extragalactic Database (NED) which is operated by the Jet Propulsion Laboratory, California Institute of Technology, under contract with NASA. 


\software{Dolphot \citep{2000PASP..112.1383D}, DrizzlePac, Matlab}


\end{document}